\documentclass[11pt]{article}
\def\be {\begin{equation}}
\def\ee {\end{equation}}
\def\bea {\begin{eqnarray}}
\def\eea {\end{eqnarray}}

\begin{document}
\title{{\bf{\Large Black hole spectroscopy via adiabatic invariance}}}
\author{
   {\bf {\normalsize Bibhas Ranjan Majhi}$
$\thanks{E-mail: bibhas@iucaa.ernet.in}}\\
 {\normalsize IUCAA, Post Bag 4, Ganeshkhind,}
 \\{\normalsize Pune University Campus, Pune 411 007, India}\\\\
 {\bf {\normalsize Elias C. Vagenas}$
$\thanks{E-mail: evagenas@academyofathens.gr}}\\
 {\normalsize Research Center for Astronomy and Applied Mathematics,}
 \\{\normalsize Academy of Athens,}
\\{\normalsize Soranou Efessiou 4, GR-11527, Athens, Greece}
\\[0.3cm]
}

\maketitle
\begin{abstract}
\par\noindent
During the last years, one had to combine the proposal about how
quasinormal frequencies are related with black holes and the
proposal about the adiabatic invariance of black holes in order to
derive the quantized entropy spectrum and its minimum change for
several black holes. In this letter we exclusively utilize the
statement that the black hole horizon area is an adiabatic
invariant and derive an equally spaced entropy spectrum of a black
hole with its quantum to be equal to the one given by Bekenstein.
Interestingly, in our approach no concept of quasi-normal mode is
needed.
\end{abstract}

\vskip 9mm

\par\noindent
It has been almost forty years since Bekenstein showed that black hole entropy is proportional
to its horizon area as well as that the horizon area is quantized in units of $l_{p}^{2}$ \cite{bekenstein1}
\footnote{It is noteworthy that the quantization of the black hole horizon in units of $l_{p}^{2}$
was also later confirmed by several different approaches (see for instance \cite{Mukhanov:1986me,Louko:1996md,Hod:1999nb}).}.
Additionally, Bekenstein proved that the quantum of the black hole horizon is given as \cite{bekenstein2}
\be
(\Delta A)_{min} = 8\pi l_{p}^{2}~.
\label{bekensteinarea}
\ee
Futhermore, Bekenstein showed that the black hole horizon area can
be considered as an adiabatic invariant
\cite{bekenstein3,bekenstein4,bekenstein5,Bekenstein:1998af}
\footnote{Following a different analysis, Barvisky, Das, and Kunstatter also showed
that the black hole horizon area is an adiabatic invariant
\cite{Barvinsky:2000gf,Barvinsky:2001tw,Barvinsky:2002qu}.}. Later
on, Hod proposed that if one employs Bohr's correspondence
principle, then the real part of quasinormal frequencies of a
black hole uniquely fixes the quantum of the black hole area
spectrum \cite{hod,Hod:2000it}. In 2002, Kunstatter combined the
proposal by Bekenstein for the adiabaticity of the black hole
horizon area and Hod's proposal for the relation between the
quasinormal frequencies, and proved that the entropy spectrum of a
d-dimensional black hole is quantized and derived the quantum of
the horizon area which is proportional to the logarithm of a
yet-to-be defined integer \cite{kunstatter}. The specific result
for the horizon area quantum is in agreement with that given by
Hod \cite{hod} as well as by Bekenstein and Mukhanov
\cite{mukhanov}. It should be noted that since the area quantum is
proportional to the logarithm of an undefined integer, one can
give a statistical description to the black hole entropy and hence
it is expected the integer to be specified by the yet-to-be found
underlying microscopic theory\footnote{In the framework of Loop
Quantum Gravity, several attempts to compute the undefined integer
have been performed \cite{corichi}.}.
In 2007, Maggiore provided a new interpretation of the black hole
quasinormal frequencies in connection to the quantum of black hole
horizon area \cite{maggiore}. In particular, his proposal was
based on the statement that perturbed Schwarzschild black holes have to be
regarded as damped harmonic oscillators whose frequencies are complex, i.e. contain both
real and imaginary parts. Thus, one should now be
interested in the absolute values of the complex quasinormal mode frequencies
while in Hod's work one was interested only for the real parts of the frequencies. In addition, Maggiore stated
that the most interesting case is that of highly excited quasinormal modes for which the imaginary part is dominant
compared to the real part. Following this new interpretation for the black hole quasinormal
frequencies and the black hole property of adiabaticity, it was later
showed that the entropy spectrum of a Kerr black hole
\footnote{The quasinormal frequencies of a Kerr black hole have been derived analytically \cite{Keshet:2007nv,Keshet:2007be}
as well as numerically \cite{Berti:2004um}.}
is also quantized evenly and the quantum of the Kerr black hole horizon area is exactly the same
with the one derived by Bekenstein
\cite{vagenas,medved} \footnote{For other approaches towards this
direction, see \cite{Banerjee:2009pf,Majhi:2009xh,Banerjee:2010be,Ropotenko:2009gr}.}.
\par\noindent
In this work we discuss the entropy spectrum and the horizon area quantum by proposing a new
approach. In particular, we derive the entropy spectrum and the horizon area quantum utilizing solely the adiabaticity of black holes.
Then using the Bohr-Sommerfeld quantization rule the entropy spectrum is derived. It is noteworthy that
there is no use at all of the quasinormal frequencies to obtain our findings. The resulting black hole entropy spectrum is equidistant
and the corresponding horizon area quantum is identical to Bekenstein's result (see equation (\ref{bekensteinarea})).\\
\par\noindent
We start by considering an  adiabatic invariant quantity of the form
\begin{eqnarray}
\int p_{i}dq_i=\int \int_0^{p_{i}} dp'_{i} dq_i=\int\int_0^H \frac{dH'}{\dot{q_i}}dq_i
\label{new1}
\end{eqnarray}
where $p_i$ is the conjugate momentum of the coordinate $q_i$ with $i=0,1$ for which  $q_0 = \tau$ and $q_1 = r$.
Here, $\tau$ is the Euclidean time and we have adopted the Einstein summation convention.
In addition, for the last step of equation (\ref{new1}) we have
implemented Hamilton's equation $\dot{q_i}=\frac{dH}{dp_{i}}$ where $H$ stands for the Hamiltonian of the system under study.
For the specific case of a black hole, the Hamiltonian $H$  is the total energy of the black hole.
\par\noindent
Then we rewrite the expression of equation (\ref{new1}) in terms of $\tau$ and $r$ coordinates
\begin{eqnarray}
\int p_{i}dq_i= \int\int_0^H \frac{dH'}{\dot{r}}dr+\int\int_0^H {dH'}d\tau~.
\label{1.1}
\end{eqnarray}
Now in order to calculate the adiabatic invariant quantity as given in equation (\ref{1.1}) for a black hole,
let us consider the general class of a static, spherically symmetric spacetime of the form
\begin{eqnarray}
ds^2 = -f(r)dt^2+\frac{dr^2}{g(r)}+r^2 d\Omega^2
\label{1.01}
\end{eqnarray}
where the location of the black hole horizon, namely  $r=r_h$, is given by equation $f(r_h)=g(r_h)=0$.
To obtain the quantity $\dot r$ that appears in equation (\ref{1.1}) we consider the radial null paths \footnote{
It should be mentioned that a black hole and a particle are {\it a priori} distinct
systems. Adiabatic invariants are quantities which vary slowly compared to changes due to external perturbations.
In our analysis, the external influence that induces changes to the black hole but leaves the horizon area unaffected
is the particle \cite{Mayo:1998ah}.}.
Since in equation (\ref{1.1}) $\tau$ stands for the Euclidean time, we first Euclideanize the metric,
given in equation (\ref{1.01}), by using the transformation $t\rightarrow -i\tau$. This leads to
\begin{eqnarray}
ds^2 = f(r)d\tau^2 + \frac{dr^2}{g(r)}+r^2d\Omega^2~.
\label{Eucliden}
\end{eqnarray}
%
%
%
%
%
The radial null paths ($ds^2=d\Omega^2=0$) are now written as
\begin{eqnarray}
\dot{r}\equiv\frac{dr}{d\tau}=\pm i \sqrt{f(r)g(r)}\equiv R_{\pm}(r)
\label{1.03}
\end{eqnarray}
where the positive (negative) sign denotes the outgoing (incoming) radial null paths.
Henceforth, our subsequent analysis will focus on the outgoing paths, since these are the ones related
to the quantum mechanically nontrivial features under consideration \cite{Parikh:1999mf}.

%
%
%
%
%
%
%
%
%
\par\noindent
Now, we employ equation (\ref{1.03}) and  it is straightforward to show that
\begin{eqnarray}
\int\int_0^H {dH'}d\tau = \int\int_0^H {dH'}\frac{dr}{R_{+}(r)}=\int\int_0^H {dH'}\frac{dr}{\dot{r}}
\label{rev1}
\end{eqnarray}
and thus the adiabatic invariant quantity given by equation (\ref{1.1})  now reads
\begin{eqnarray}
\int p_{q_i}dq_i= 2\int\int_0^H {dH'}d\tau ~.
\label{1.4}
\end{eqnarray}
In order to perform the $\tau$-integration, we first note that $\tau$ has periodicity
$\frac{2\pi}{\kappa}$, where $\kappa$ is the surface gravity of the black hole
\begin{equation}
\kappa = \frac{1}{2}\sqrt{f'(r_h)g'(r_h)}~.
\end{equation}
However, since we are considering only the outgoing paths the integration limit for
$\tau$ will be $0\leq\tau\leq\frac{\pi}{\kappa}$ and hence the adiabatic invariant quantity reads
\begin{eqnarray}
\int p_{q_i}dq_i = 2\pi\int_{0}^{H} \frac{dH'}{\kappa}~.
\label{1.5}
\end{eqnarray}
It is known that the temperature of a black hole is proportional to the surface gravity of the horizon
\begin{eqnarray}
T_{bh}=\frac{\hbar\kappa}{2\pi}
\label{1.6}
\end{eqnarray}
and thus the adiabatic invariant quantity given in equation (\ref{1.5}) becomes
\begin{eqnarray}
\int p_{q{_i}}dq_i=\hbar\int_{0}^{H} \frac{dH'}{T_{bh}}=\hbar S_{bh}
\label{1.7}
\end{eqnarray}
where in the last step we have exploited the first law of black hole thermodynamics
\be
\int_0^H\frac{dH}{T_{bh}}=S_{bh}~.
\ee
%

%
%
%
%
%
\par\noindent
Finally, implementing the Bohr-Sommerfield quantization rule
\begin{eqnarray}
\int p_{q_{i}}dq_i=nh
\label{1.10}
\end{eqnarray}
%
%
%
in equation (\ref{1.7}), we derive the black hole entropy spectrum \footnote{From a different perspective,
this spectrum was also derived from Barvinsky and Kunstatter,
and, independently, from Kastrup \cite{Kastrup:1996pu,Barvinsky:1996hr,Barvinsky:1996bn}.}
\begin{eqnarray}
S_{bh}=2\pi n \hspace{3ex} \mbox{where} \hspace{1ex}\ n=1,2,3,\ldots
\label{1.11}
\end{eqnarray}
\par\noindent
and it is evident that the spacing in the entropy spectrum is given by
\begin{eqnarray}
\Delta S_{bh}=S_{(n+1)}{_{bh}}-S_{(n)}{_{bh}}=2\pi ~.
\label{1.12}
\end{eqnarray}
Therefore, the entropy spectrum is quantized and equidistant for a spherically symmetric static black hole.
\newline
\par\noindent
At this point some comments are in order.
Firstly, in the framework of Einstein's theory of gravity, black hole entropy is proportional to
the black hole horizon area \cite{bekenstein2}
\begin{eqnarray}
S_{bh}=\frac{A}{4l_p^2}~.
\label{1.15}
\end{eqnarray}
It is evident that if we employ the spacing of the entropy spectrum given in equation (\ref{1.12}),
the quantum of the black hole horizon area is of the form
\begin{eqnarray}
\Delta A=8\pi l_p^2 ~.
\label{1.16}
\end{eqnarray}
This is identical to the area quantum derived by Bekenstein (see equation (\ref{bekensteinarea})).
\par\noindent
Secondly, it should be stressed that although the
present analysis was developed in a static and spherically symmetric black hole spacetime,
the entropy/area quantization of a rotating black hole can also be
discussed in this context. It is well known that the near-horizon
theory of a black hole is mainly governed by a ($1+1$)-dimensional
metric whose form is just the ($t-r$)-sector of the metric given in equation (\ref{1.01})
\footnote{A detailed analysis for the case of Kerr-Newman black hole
is given in \cite{Umetsu:2009ra}.}. Since the  radial null path depends only on the
($t-r$)-sector, it is evident that for the rotating case one will find exactly the same
path with that given in  equation (\ref{1.03}). The only constraint here is that we are working
in the near-horizon regime. However, this is not a problem since the interesting quantum black hole phenomena are
those related to near-horizon effects. Thus, up to equation (\ref{1.12}) the approach
will be the same for the case of a rotating black hole.
Therefore, the spacing in the entropy spectrum of a Kerr black hole will be
equal to the one given in equation (\ref{1.12}).
\par\noindent
Thirdly, it may be interesting to give a connective discussion between the work of Bekenstein \cite{Bekenstein:1998af}
and our present work. Bekenstein proved that if a particle of a specific relativistic energy moves along the radial path
and is absorbed by a non-extremal black hole from the classical turning point ($\frac{dr}{d\tau} =0$),
the horizon area remains invariant and hence it is an adiabatic invariant quantity. On the other hand,
if the energy is larger than the specific value of the relativistic energy, the area cannot be treated as an adiabatic invariant quantity.
Therefore, one can always hope to discuss the quantization of black hole in the spirit of the old quantum theory.
Here we discussed the entropy/area quantization utilizing this spirit. We started with a formal classical
adiabatic invariant quantity. This consisted of two parts which were shown to be equal by using the outgoing
radial path. The explicit calculation revealed that the quantity is proportional to the entropy of the black hole.
Then using the Bohr-Sommerfeld quantization rule the spectra of entropy and area were obtained.
\par\noindent
Lastly, in more general theories of gravity that lay beyond the regime of validity
of Einstein's gravity, entropy is a function of the area of the black hole \cite{paddy,Wei:2009yj}
\begin{eqnarray}
A= F(S_{bh})
\label{1.13}
\end{eqnarray}
and the corresponding quantum of the black hole horizon area now reads
\begin{eqnarray}
\Delta A =\frac{\partial F}{\partial S_{bh}}\Delta S_{bh}=2\pi \frac{\partial F}{\partial S_{bh}}~.
\label{1.14}
\end{eqnarray}
If the horizon area, $A$, is not proportional to black hole entropy, $S_{bh}$, as it is in the case of Einstein's gravity,
then the quantum of the horizon area is not a constant and so the horizon area spectrum is not evenly spaced
though the entropy spectrum is. As an example, we consider the black hole entropy of the form
\begin{eqnarray}
S_{bh}=\frac{A}{4l_p^2}+\alpha\ln A
\label{1.2}
\end{eqnarray}
where the constant $\alpha$ is equal to $-\frac{3}{2}$ in the context of Loop Quantum Gravity \cite{kaul},
or $\alpha$  is related to the trace anomaly \cite{fursaev,bibhas}.
The corresponding quantum of the horizon area will be of the form
\be
\Delta A=\frac{\partial A}{\partial S_{bh}}\Delta S_{bh}=\frac{8\pi l_p^2}{1+\frac{4l_p^2}{A}}\simeq8\pi l_p^2(1-\frac{4\l_p^2}{A})
\ee
and it is easily seen that the black hole horizon area spectrum is not evenly spaced.
\newline
\par\noindent
To conclude, we have derived the entropy spectrum for the more general class of spherically symmetric static black holes
as well as its corresponding horizon area quantum by proposing a new approach in the context of adiabatic invariant quantities.
In particular, Hod's proposal (or Maggiore's proposal) about black hole quasinormal frequencies has not been
employed in this analysis. It is noteworthy that the obtained black hole entropy spectrum is evenly spaced while the quantum
of the horizon area is exactly the same with Bekenstein's result.
The approach is general enough to include other types of black hole spacetimes.\\

\par\noindent
{\bf Acknowledgments}\\
We are grateful to G. Kunstatter and J.M.~Pons for useful discussions and enlightening comments.


\end{document}